# Lineshapes in Nuclear Forward Scattering Frequency Spectra

Alexandre I. Rykov

*School of Engineering, The University of Tokyo, Japan*

*and Siberian Synchrotron Radiation Centre, Novosibirsk, Lavrentieva 11, 630090 Russia*

**Abstract**: A new method for investigating the particle size distribution (PSD) is suggested using the nuclear forward scattering (NFS) of synchrotron radiation. The distribution of Weibull-Gnedenko is used to illustrate the relationship between the NFS lineshape and PSD.

Nuclear exciton polariton [1,2] manifests itself in the lineshape of the spectra of nuclear forward scattering Fourier-transformed from time domain to frequency domain. This lineshape is generally described by the convolution of two intensity factors. One of them is Lorentzian related to free decay. We derived the expressions for the second factor related to Frenkel exciton polariton effects at propagation of synchrotron radiation in resonant media. Parameters of this Frenkelian shape depend on the spatial configuration of the Mössbauer media. In a layer of thickness, distributed exponentially, we found this shape having the following functional form (Fig. 1)

$$I(\omega) = L_a \left[ 1 - \frac{\sqrt{\pi} \cos\left(\tfrac{1}{2} \arctan\left(4 L_a / \pi \Delta\omega t_0\right)\right)}{(\pi^2 + 16 L_a^2 / \Delta\omega^2 t_0^2)^{1/4}} \right] \quad (1)$$

Here $L_a$ is the distribution-averaged chord length expressed in units of dimensionless Mössbauer thickness ¼ $\pi Z \sigma_0 \rho f \eta$, Z is the linear length in cm, $\sigma_0$ = 2.56·10⁻¹⁸ cm² is resonant cross-section, $\rho$ is the density of the resonant nuclei, f is the Lamb-Mössbauer factor, and $\eta$ is the resonant nuclide isotope abundance.

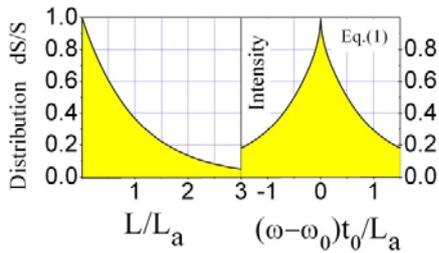

Fig. 1. Absorber thickness (left) and spectral intensity (right) distributions. Here dS/S is the relative area of absorber having thickness L. Angular frequency $\omega$ is in rad/s and the free decay time $t_0$ is in seconds, e.g, for ⁵⁷Fe $t_0$=1.41·10⁻⁷ s. Here $\Delta\omega = \omega-\omega_0$ and $\omega_0$ is the frequency of quantum beats.

The Eq. (1) is obtained via Fourier transform $\Phi(f) = \sqrt{2/\pi} \int_0^\infty f(t) \cos(\omega t) dt$ of the squared Bessel function of first kind and order one $J_1$ integrated with the exponential distribution of the radiation pathway lengths:

$$I(\omega) = \Phi \left[ \frac{\zeta}{\sqrt{2\pi}} \frac{e^{-\frac{\zeta}{L_a}}}{L_a} \left( \frac{J_1\left(2\sqrt{\zeta t/\pi t_0}\right)}{\sqrt{\zeta t/\pi t_0}} \right)^2 \right]$$

The Eq.(1) is of great importance because the materials with exponential chord lengths distribution are widespread in the nature [3].

In powdered materials, we show that when the particle size distribution (PSD) is described by the density function of Weibull-Gnedenko (WG)

$$F_{WG}(r) = \frac{c}{R}\left(\frac{r}{R}\right)^{c-1} \exp\left[-\left(\frac{r}{R}\right)^c\right], \qquad (2)$$

then the distribution of chords $\zeta$ in a layer of spherical particles is described by

$$d_{WG}(\zeta) = \left[\Gamma\left(\frac{2+c}{c}\right)\right]^{-1} \frac{\zeta}{2R^2} \exp\left[-\left(\frac{\zeta}{2R}\right)^c\right] \qquad (3)$$

Here $c$ is the shape parameter, $R$ is the scale parameter and $\Gamma$ denotes the gamma function. The WG distribution is very convenient for the parametrization of wide range of PSD's, including exponential (c=1) and Rayleigh (c=2) distributions.

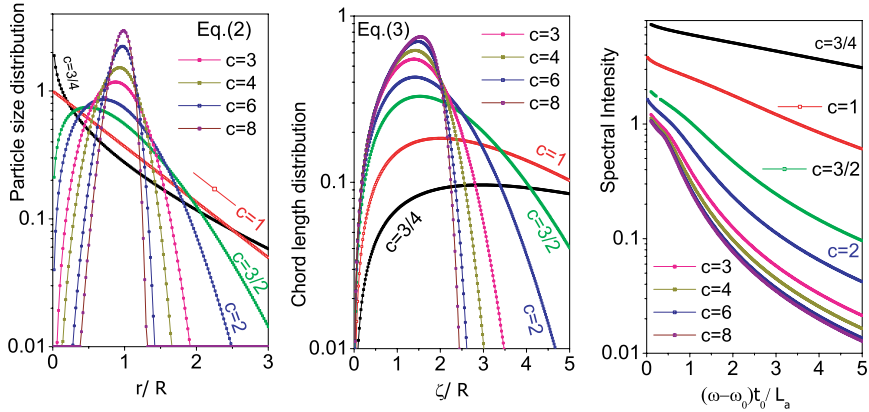

Fig. 2. Distributions of Weibull-Gnedenko for particle sizes, chord lengths and spectral intensities. The spectra c=1 and c=2 in right-hand panel correspond to Eqs. (5) and (6), respectively.

The Eq.(3) was derived from Eq.(2) using the expressions for CLD of spheres and second moment of the distribution $\langle r^2 \rangle$:

$$d(\zeta_n) = \frac{\zeta}{2\langle r^2 \rangle} \int_{\zeta/2}^{\infty} F(r)dr \qquad \text{with} \qquad \langle r^2 \rangle = R^2 \Gamma\left(\frac{c+2}{c}\right) \qquad (4)$$

In the NFS frequency spectra, the intensity for particles with exponential PSD (c=1) is

$$I_1(\omega) = 4R\left[1 + \frac{2\sqrt{\pi}R\sin\left[\frac{3}{2}\arctan(\frac{8R}{\pi\Delta\omega t_0})\right]}{\Delta\omega t_0(\pi^2 + 64R^2/\Delta\omega^2 t_0^2)^{3/4}} - \frac{\sqrt{\pi}R\cos\left[\frac{1}{2}\arctan(\frac{8R}{\pi\Delta\omega t_0})\right]}{(\pi^2 + 64R^2/\Delta\omega^2 t_0^2)^{1/4}}\right] \qquad (5)$$

The Rayleigh PSD (c=2) results in the hypergeometric spectral lines:

$$I_2(\omega) = \sqrt{\pi}R\left[1 - {}_2F_2\left(\{\frac{1}{3},\frac{3}{4}\},\{\frac{1}{2},1\},-\frac{16R^2}{\pi\Delta\omega^2 t_0^2}\right)\right] \qquad (6)$$

Note that the shoulder in the intensity profile becomes less pronounced and shifts to higher $\Delta\omega$ as the shape parameter c decreases. The shoulder is still well discernible at $\Delta\omega\, t_0 \sim 2L_a$ for the exponential (c=1) PSD. On the other hand, in case of exponential CLD (Eq.1), the intensity profile is fully concave without any discernible shoulder.

In conclusion, the direct and inverse scattering problems are tackled to find the spatial configuration of resonant media. We have solved the direct problem of determination of the NFS spectral lineshape starting from several characteristic size distributions of particles. Inverse problem of finding the characteristic parameters of particle chord length distribution from polaritonic lineshapes would be possible to solve using regularization methods. This method requires preparing the single layer of particles spread over 2D surface. Another way of sample preparation may involve making an absorber of Mössbauer particles well diluted in non-Mössbauer media [4] to avoid shadowing of one particle by another. Then, not only mean particle size can be found from characteristic linewidth, but also parameters of particle size distribution can be found from the lineshape.